\documentclass{PoS}
\usepackage[authoryear]{natbib}
\bibpunct{(}{)}{;}{a}{}{,}

\title{Overview of Complementarity and Synergy with Other Wavelengths in Cosmology in the SKA era}

\ShortTitle{Complementarity and Synergy in Cosmology}

\author{
\speaker{Keitaro Takahashi}$^1$,
Michael L. Brown$^2$,
Carlo Burigana$^{3,4}$,
Carole A. Jackson$^5$,
Matt Jarvis$^6$,
Thomas D. Kitching$^7$,
Jean-Paul Kneib$^8$,
Masamune Oguri$^9$,
Simon Prunet$^{10}$,
Huanyuan Shan$^8$,
Jean-Luc Starck$^{11}$,
Daisuke Yamauchi$^{12}$
\\
$^1$Graduate School of Science and Technology, Kumamoto University, Japan
$^2$Jodrell Bank Centre for Astrophysics, University of Manchester, Oxford Road, Manchester M139PL, UK
$^3$INAF-IASF Bologna, Via Piero Gobetti 101, I-40129, Bologna, Italy
$^4$Dipartimento di Fisica e Scienze della Terra, Universit\`a degli Studi di Ferrara, Via Giuseppe Saragat 1, I-44100 Ferrara, Italy
$^5$ICRAR, Curtin University, GPO Box U1987, Perth 6845, Australia
$^6$Astrophysics Department, University of Oxford, UK
$^7$Mullard Space Science Laboratory, University College London, Holmbury St Mary, Dorking, Surrey RH5 6NT, UK
$^8$Ecole Polytechnique F\'ed\'erale de Lausanne Laboratoire d'Astrophysique, Observatoire de Sauverny, CH-1290 Versoix, Switzerland
$^9$Graduate School of Science, University of Tokyo, Japan
$^{10}$Canada-France-Hawaii Telescope
$^{11}$Laboratoire AIM, UMR CEA-CNRS-Paris 7, Irfu, SAp, CEA Saclay, F-91191 Gif sur Yvette cedex, France
$^{12}$Research Center for the Early Universe, University of Tokyo, Japan
\\
E-mail:\email{keitaro@sci.kumamoto-u.ac.jp}
}

\abstract{
We give an overview of complementarity and synergy in cosmology between the Square Kilometre Array and future survey projects in other wavelengths. In the SKA era, precision cosmology will be limited by systematic errors and cosmic variance, rather than statistical errors. However, combining and/or cross-correlating multi-wavelength data, from the SKA to the cosmic microwave background, optical/infrared and X-ray, substantially reduce these limiting factors. In this chapter, we summarize future survey projects and show highlights of complementarity and synergy, which can be very powerful to probe major cosmological problems such as dark energy, modified gravity and primordial non-Gaussianity.
}

\FullConference{
Advancing Astrophysics with the Square Kilometre Array\\
June 8-13, 2014\\
Giardini Naxos, Italy}

\newcommand{\skipthis}[1]{}

\begin{document}

\section{Introduction}

In the era of SKA operations, astronomy will be fully in the realm of ebig dataf science. Observational data will have been accumulated from millions of discrete objects across the electromagnetic spectrum by a large number of ground and space-based missions. In addition there will be existing and evolving detailed  simulated data: all of which the SKA will contribute to and provide a unique insight to its interpretation.

Whilst significant sets of data may be complete from a number of other facilities ahead of, and during early SKA operations, the challenge will be to fully exploit its scientific yield. This requires that observational integrity and repeatability are maintained across the build phases to realize the full SKA (SKA2). This alignment of the data is critical because many SKA-complementary facilities will only have limited lifetimes, and as such it relies on the \emph{SKA} meeting these requirements not vice-versa. It should also be noted that the SKA will come into operation in an era where we will have moved beyond multiple simultaneous space missions as we do now due to increasing costs. Even today, some of the SKA-complementary facilities are of limited-lifetime (even single-experiment) facilities. 

So far cosmology has been developed by extracting statistical information from the distribution of a large number of objects such as galaxies and galaxy clusters. Errors in cosmological parameters are, in many cases, dominated by statistical errors, which can be reduced by increasing the number of objects. However, in the SKA era with huge optical/infrared surveys by Euclid, LSST and WFIRST, systematic errors and cosmic variance, rather than statistical errors, will become the limiting factors to advance cosmology further. Thus, it is of vital importance to reduce systematics and cosmic variance.

Instrumental systematics will be independent for different wavelength observations so that combining or cross-correlating multi-wavelength data is expected to be able to reduce systematics substantially. This is a major way of synergy and weak lensing is a typical example as we describe briefly later. Cross-correlation is also effective between the CMB anisotropies and large scale structure. Although the CMB anisotropies contain a lot of cosmological information such as initial condition and evolution of primordial fluctuations, geometry of the universe and history of cosmic expansion, it is not easy to disentangle them. Cross-correlation with the distribution of galaxy is a powerful way to extract low-redshift information, which allows us to investigate dark energy and modified gravity.

In this chapter, we give an overview of synergies in cosmology between the SKA and surveys in other wavelengths. Each topic will be briefly described and we refer to other chapters in  for the details.

\section{Future Survey Projects}

In this section we briefly summarise the major surveys that are, or will, take place until routine SKA1 operations. In writing about the future potential, we have to bear in mind that timelines can slip and missions change in scope and focus; however what follows is a summary of the cosmology-relevant landscape that SKA will inhabit. We must also consider SKA as it evolves: the initial build phase (SKA1) will have a limited operational window, followed by potential build interruptions to evolve into the full SKA (SKA2).

\subsection{Surveys in the pre-construction SKA era}

There are many SKA pathfinding surveys during the pre-construction and early build era; these use existing and precursor/pathfinder telescopes. Many of these are major wide-field and/or deep surveys in themselves. They will deliver important results as well as informing both the design and science directions of the SKA. These surveys are discussed in Norris et al. (2014) and their details are not presented here.  A characteristic of these pathfinding surveys is the preparatory work being undertaken by large international teams ahead of observations. Not only will this work tackle some of the data management and SKA-era compatibility issues but it will also ensure the systematics of these surveys is well understood, ready for even deeper cosmological analyses. 

Across the rest of the spectrum there are a number of major new facilities that will yield data to progress SKA-cosmology experiments. In the optical these include SDSS, JWST, DESI and VST/KiDS. In the infrared WISE and HERSCHEL have already contributed massive information which SKA observations will draw on. At higher energies eROSITA (X-ray) will detect clusters as a probe of dark energy and primordial density fluctuations \citep{colafrancesco}  and GALEX (UV) are expected to have completed observations by the first operations of SKA.

\subsection{Surveys in the SKA era $\sim$ 2020 onwards}

At the time that the SKA is operational, there will be additional survey data flowing from a number of new ground-based telescopes and space missions that we now briefly discuss.

Euclid \citep{Euclid} is expected to be launched in 2020 and will perform imaging and spectroscopic surveys in optical and infrared bands. The survey area is about 15,000 ${\rm deg}^2$ and the primary sciences are weak lensing and baryon acoustic oscillation. The synergies between the SKA and Euclid are discussed in the other chapter \citep{kitching}.

The large synoptic survey telescope (LSST) \citep{LSST} is an ground-based optical wide-field survey telescope that will observe the entire available sky every few nights and anticipated to be operational from 2022. It will observe 18000 square degrees of the southern hemisphere and provide photometric redshifts and optical shapes. The synergies between the SKA and LSST are discussed in the other chapter \citep{bacon}.

Great efforts are being pursued in the CMB community to prepare a next generation of experiments. These are space missions and include the following.

In response to the ESA Cosmic Vision (2015-2025) Call for Proposals, a medium-size mission, B-Pol (http://www.b-pol.org) has been proposed. This is targeted to ultra-accurate CMB polarization measurements up to a moderate resolution (about 1 degree) at six frequency bands between 45 GHz and 353 GHz \citep{debernardisetal09} to primarily identify primordial B-modes.  B-pol is proposed to be realized as a set of eight small telescopes co-aligned within the spacecraft axis with an array of about 1000 single mode corrugated feed horns in each telescopefs focal plane, designed to be well-matched with the optics and only minimal aberrations.   

Building on {\it Planck}'s success the Cosmic Origins Explorer (COrE: http://www.core-mission.org) \citep{corecoll} has been proposed in response to the ESA call. About 6400 dual-polarisation receivers at the focal plane of a 1.5 - 2 m class telescope would achieve a resolution of a few arcmin with excellent polarization sensitivity, across a wide (45 GHz - 795 GHz) frequency range (15 bands). The goal of COrE is to extend CMB polarization studies through to higher multipoles, probing a variety of cosmological and fundamental physics along with a providing a new generation of all-sky polarization surveys. 

In the other critical aspect of CMB cosmology, i.e. the detailed study of the frequency spectrum, little progress has been made since the stunning results from COBE/FIRAS. There are two ambitious projects now proposed which would have impact in the SKA era. The Primordial Inflation Explorer (PIXIE) \citep{kogutpixie} has been presented to NASA as an Explorer-class proposal. It would be equipped with receivers operating between 30 GHz and 6 THz, sensitive to polarization and coupled to an advanced cryogenic calibration system to achieve the required precise measurements.  The main goal is to reveal the fine details of the CMB spectrum as well as addressing the extremely small scales of primordial perturbations that are otherwise damped and unobservable in anisotropies. The second, and by far most ambitious mission, the Polarized Radiation Imaging and Spectroscopy Mission (PRISM: http://www.prism.org) \citep{prismjcap} was first proposed in 2013 as an ESA Science Programme. PRISM is aimed at ultra-accurate CMB mapping in both temperature and polarization, limited only be cosmic variance and foregrounds. PRISM would have imaging accuracy to a few arcmin over a very wide frequency range (30 GHz to 6 THz), and with these capabilities it would map the distribution of galaxy clusters through to the IR. Furthermore, PRISM sensitivity to CMB spectrum is about one order of magnitude better than PIXIE.

LiteBird (JAXA proposed future mission: http://litebird.jp/eng/) which will undertake a full sky CMB polarization survey at degree scale; 50 GHz - 320 GHz (with 30 arcmin resolution at 150 GHz). Planned launch in the early 2020s.

\section{Weak Lensing\label{sec:weak-lensing}}

\subsection{Weak Lensing survey efficiency}

Weak lensing is made possible by the statistical study of the shapes of distant
sources and brings information on the mass distribution located between us and these distant sources. The weak lensing measurements thus requires shape information and redshift information. Shape information is essential, however for some application redshift information can be used on a statistical basis.

The weak lensing information will scale with the number density of distant sources, and is limited by the intrinsic ellipticity distribution of the background sources and our (limited) knowledge of the PSF and its possible variation across the survey. This can be seen in the shear measurement error that can be written as:
\begin{equation}
\sigma^{2}_{\gamma}\propto {\sigma^{2}_{\varepsilon-int}+\sigma^{2}_{meas} \over N}
\end{equation}
where $N$ is the number of galaxies over which the shape is measured, $\sigma_{\varepsilon-int}$ is the intrinsic shape dispersion and depends only on the intrinsic nature of the sources observed and $\sigma_{meas}$ is the noise added by the measurement techniques and includes information (or lack of) of the PSF and the photon noise. Note that PSF issues are much more critical at optical wavelength (specially for ground based observatory, less for space mission) than radio wavelength.

Because distant sources are small (of the order of $\sim$1 arcsec or smaller), cosmological weak lensing survey efficiency will scale with the total number of sources for which one can resolve their shape. 

Different weak lensing techniques can inform us on the matter distribution on different scales.
\begin{itemize}
\item Cosmic shear allows to probe the matter power spectrum, and is very efficient in probing the very large scales. Yet, ground based weak lensing power has been quite limited on large scale ($> 100$ Mpc) because of systematic limitations.
\item Weak lensing mass mapping and peak statistics: blind search of structure, sensitive to cosmological parameters.
\item Cluster weak lensing: need cluster survey sample
\item Galaxy-galaxy lensing: need foreground galaxy sample
\end{itemize}

\begin{figure*}
\begin{center}
\includegraphics[width=0.25\textwidth]{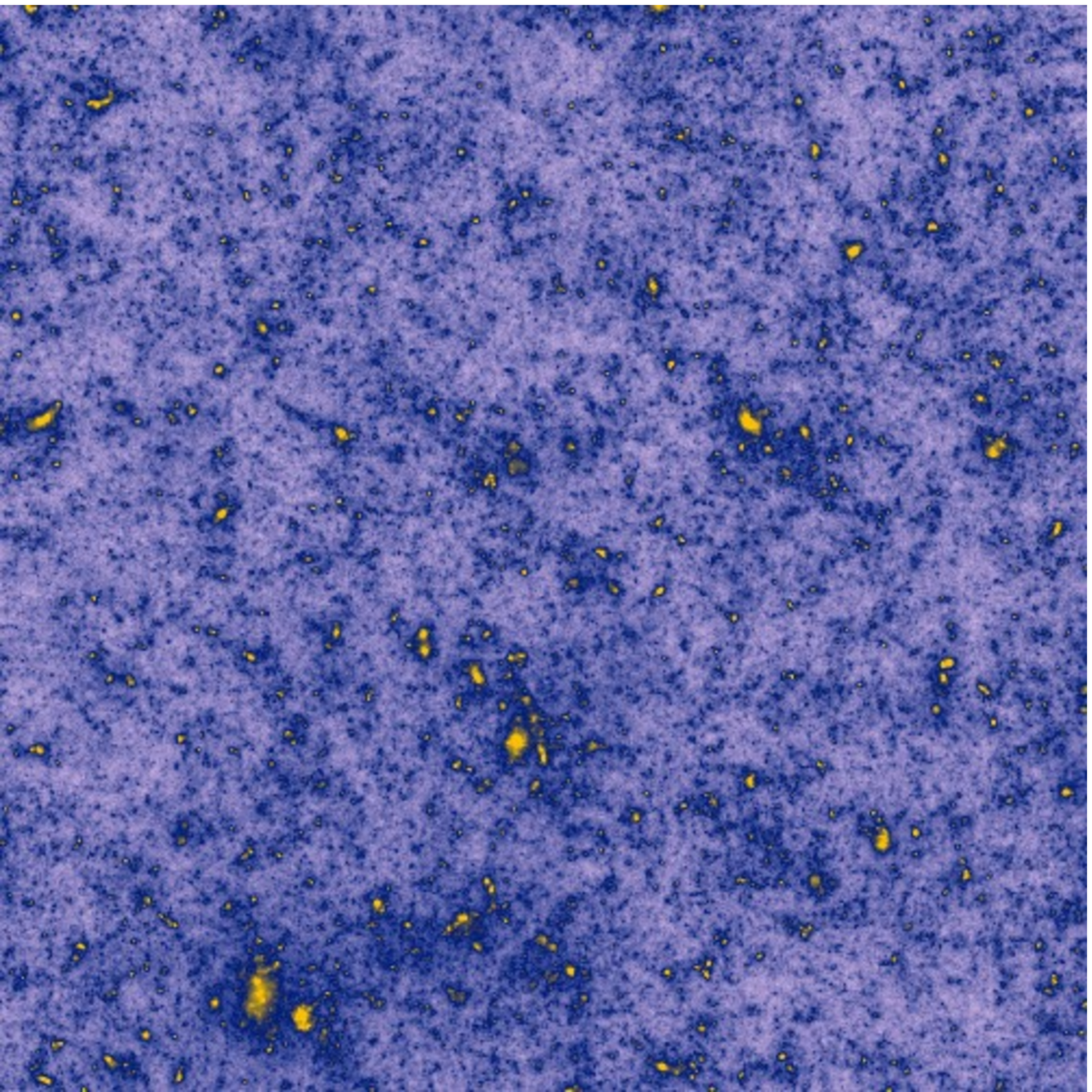}
\includegraphics[width=0.25\textwidth]{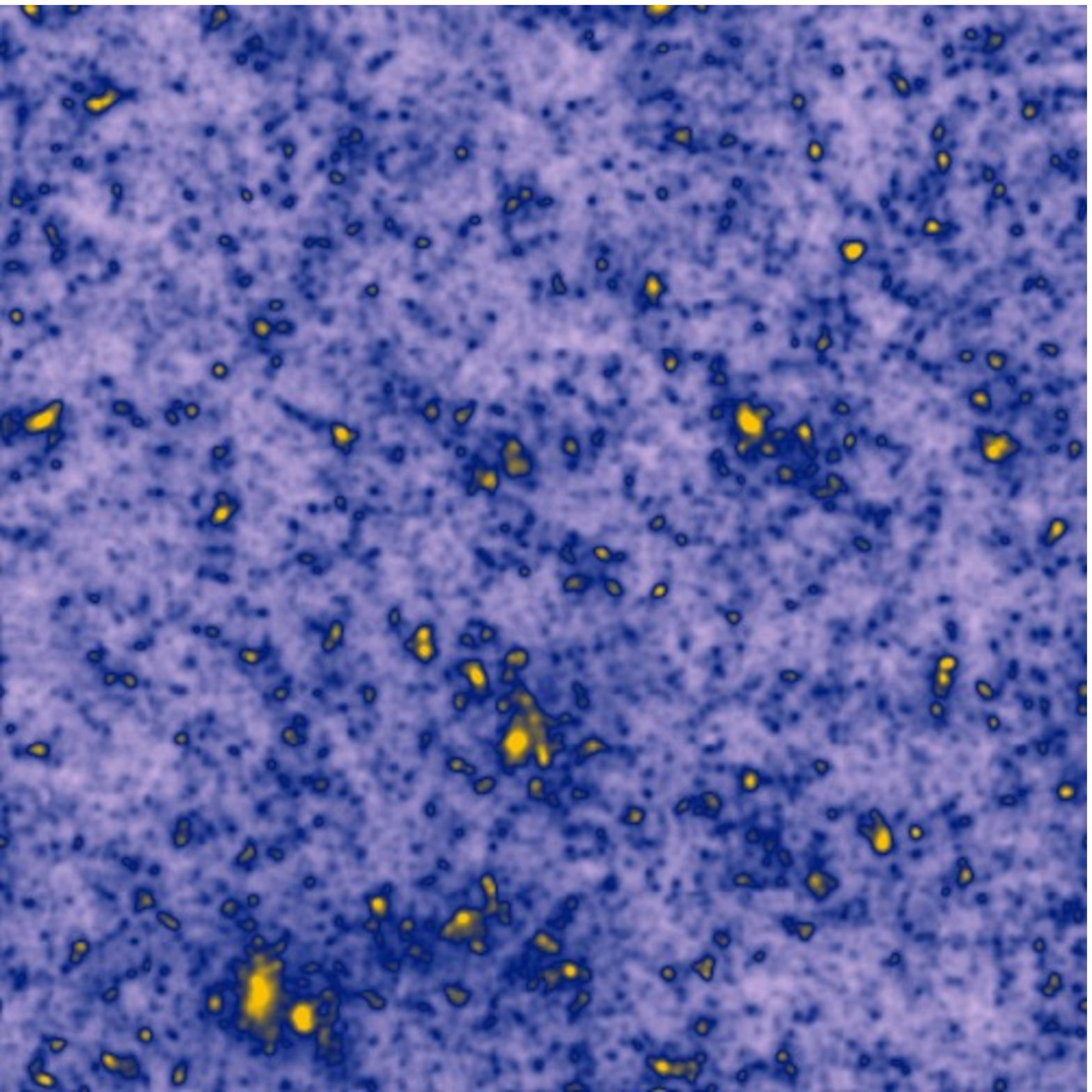}\\
\includegraphics[width=0.25\textwidth]{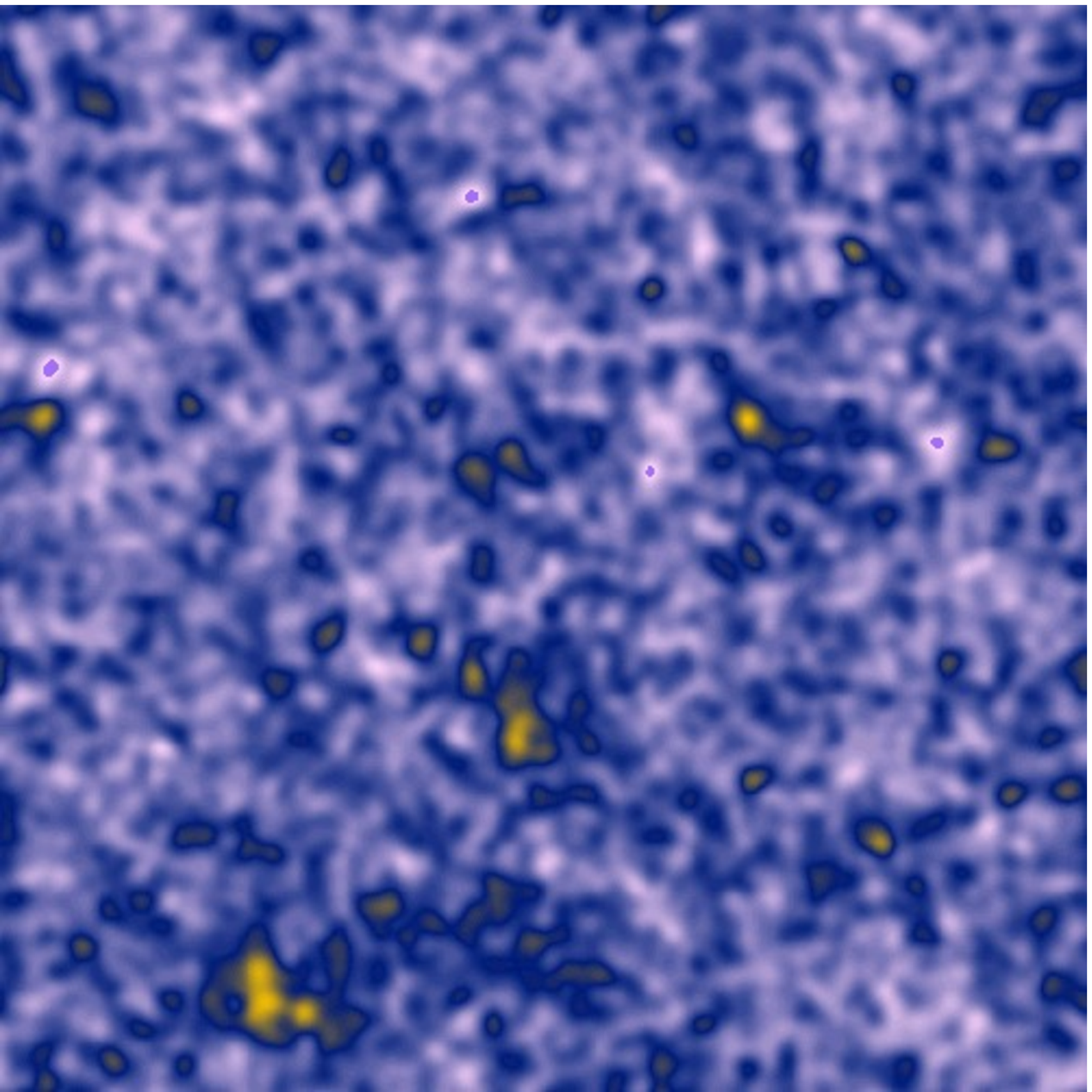}
\includegraphics[width=0.25\textwidth]{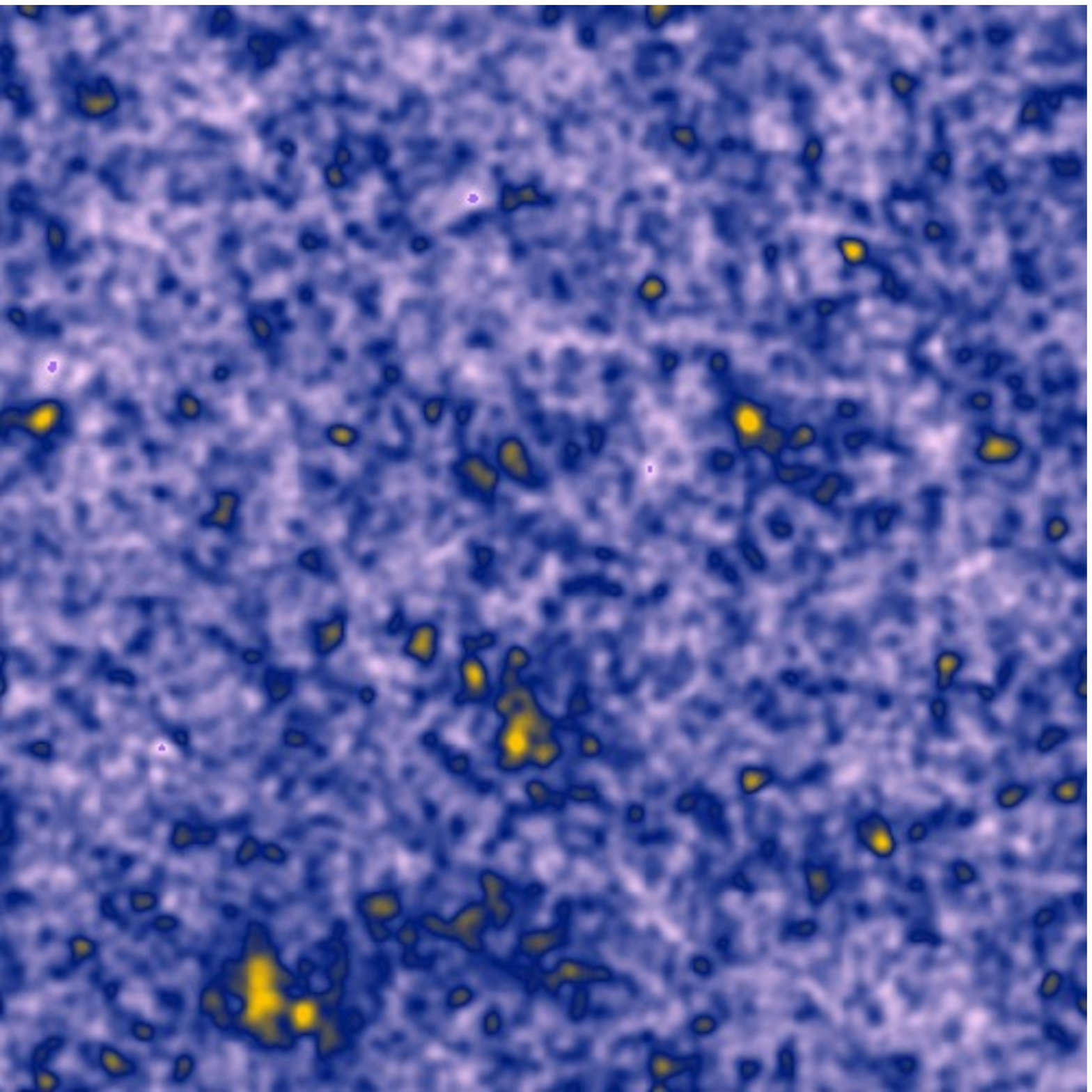}
\caption{Simulation of $3\time3$ square degrees converence maps. The top-left panel is the input mass map, the top-right panel is the smoothed mass map. The bottom-left panel shows the recovery for a Euclid-like survey with effective galaxy number density $n_g\sim 30~\rm gals/arcmin^2$. The bottom-right panel shows the simulated recovery for the SKA plus Euclid with $n_g\sim 100~\rm gals/arcmin^2$. It shows that SKA information will help increase both the weak lensing S/N measurement as well as better resolving smaller-scale mass structure.}
\label{fig:massmap}
\end{center}
\end{figure*}

\subsection{Weak lensing survey complementarity}

Wide-field radio and optical survey differs significantly by the nature of the sources found the continuum imaging. Radio sources are good tracers of star-formation activities while I-band selected galaxies used in optical are more of a tracer of stellar masses (at least to $z<1$). Radio sources are thus likely less biased system than optical sources, and possibly have a more homogeneous distribution on the sky.

Because of this, the redshift distribution of continuum radio and optical galaxies are somewhat different. Optically I-band selected sources will mostly lie at $z<1$, while radio sources are more broadly distributed in redshift with a more prominent tail at high redshift.

Radio surveys will thus allow to probe mass distribution and the large-scale structure at higher redshift than can possibly be done with optical surveys. They will also be more efficient in probing high density region such as galaxy clusters as the cluster light is blocking somewhat the  signal of background galaxies.

Overlapping optical and radio surveys have a particularly useful synergy in terms of reducing and quantifying the impact of systematic effects in weak gravitational lensing analyses (Fig. \ref{fig:massmap}). By cross-correlating the shapes of galaxies as measured in the optical and radio surveys, one can eliminate instrumental systematic effects that are not correlated between the two telescopes. Given the very different designs and modes of operation of optical and radio telescopes, one would not expect their instrumental systematic effects to be correlated and so this offers a route to measuring the cosmic shear signal in a very robust way.

Moreover, radio surveys offer unique additional ways to measure the lensing signal that are not available to optical telescopes. In particular, both radio polarization information and rotational velocity measurements from HI observations can provide estimates of the \emph{intrinsic} position angles of the lensing source galaxies. Such measurements offer great potential to (i) reduce the effects of galaxy ``shape noise'' due to the intrinsic dispersion in galaxy shapes \citep{morales} and (ii) to mitigate the contaminating signal from the intrinsic alignments in galaxy orientations which is perhaps the most worrisome astrophysical systematic effect facing future weak lensing surveys \citep{patel}. In addition to using this information in a combined analysis, one could potentially use the SKA-based estimates of the intrinsic alignment contamination to calibrate out the alignment signal in the LSST lensing survey. 

Finally, the envisaged SKA surveys will probe a wider range of redshifts than will be reached by LSST. The SKA surveys thus provide extra (high-redshift) tomographic slices with which the evolution of structre at relatively early times can be probed. SKA can push to even higher reshift by measuring the lensing distortion signal in HI intensity mapping surveys. Thus, these high-redshift SKA lensing experiments will naturally help fill the gap between the traditional optical lensing probes (where sources are typically located at $z \sim 1$) and the ultimate lensing source of the CMB at $z \sim 1000$.

\section{Cosmic Microwave Background}

The Cosmic Microwave Background is a very powerful probe to constrain cosmological parameters that are dynamically relevant at the epoch of recombination ($z\sim1080$). Recent high-sensitivity, high-angular resolution measurements of the CMB temperature anisotropies by  {\em Planck} \citep{Planck2013-1,Planck2013-15} over the nominal 14 month mission show that the now standard $\Lambda$CDM model is an excellent fit to the data. By the end of this year, the full temperature data, as well as results from polarisation data should be made public. If the quality of the CMB data is now good enough to break degeneracies between parameters that were plaguing earlier datasets, it cannot nevertheless tell us much about the precise dynamical behaviour of the Universe at low redshifts, in particular it cannot give any precise constraints on dark energy or modified gravity models. 

On the other hand, future giant radio surveys with the SKA will provide an exquisite view of the low redshift universe, with the first phase surveys at the horizon of 2020, i.e. on a similar time frame to the next generation of optical/near-infrared galaxy surveys, e.g. Euclid and LSST. These low-redshift probes however cannot constrain by themselves the full cosmological model, but they will have a tremendous impact on the study of dark-energy models \citep{Blake2004}. Therefore, the primary synergy of CMB and SKA data will come from a joint analysis of the cosmological parameters including dark energy and/or modified gravity models on linear scales where the theoretical predictions are well understood. 
\begin{figure}
\begin{center}
\includegraphics[width=0.8\textwidth]{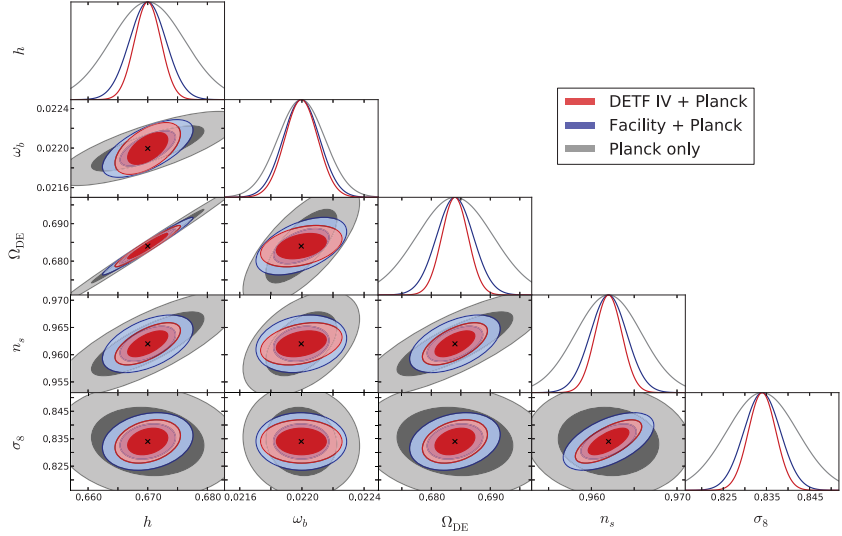}
\end{center}
\vspace{-1cm}
\caption{Compared constraints from a Fisher matrix analysis on $\Lambda$CDM parameters using different probe combinations \citep{Bull2014}. "Facility" is representative of the SKA1 survey in combined (single dish plus interferometric) mode, and "DETF IV" is representative of e.g. the Euclid redshift survey. Both SKA1 and Euclid should come on similar time frames (around 2020) and have similar power on this model. }
\label{fig:LCDM}
\end{figure}

\begin{figure}
\includegraphics[width=0.6\textwidth]{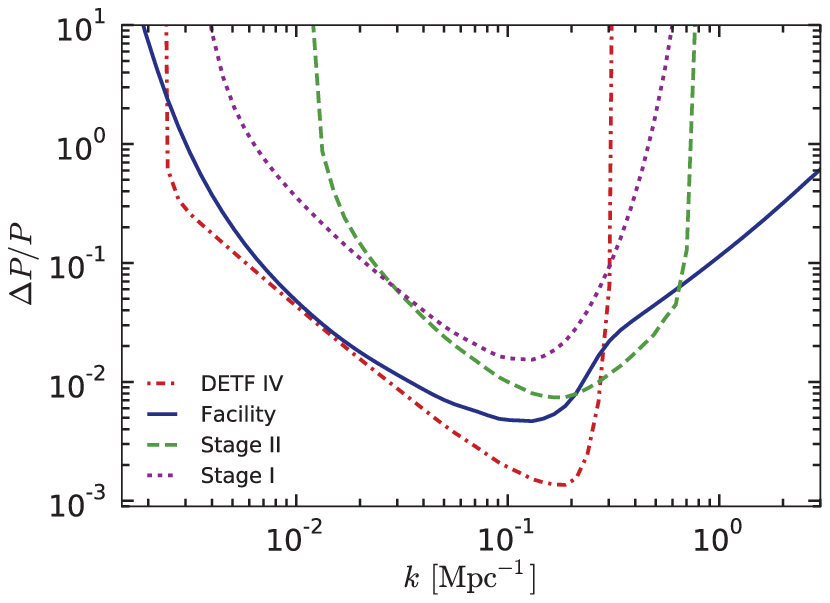}
\includegraphics[width=0.4\textwidth]{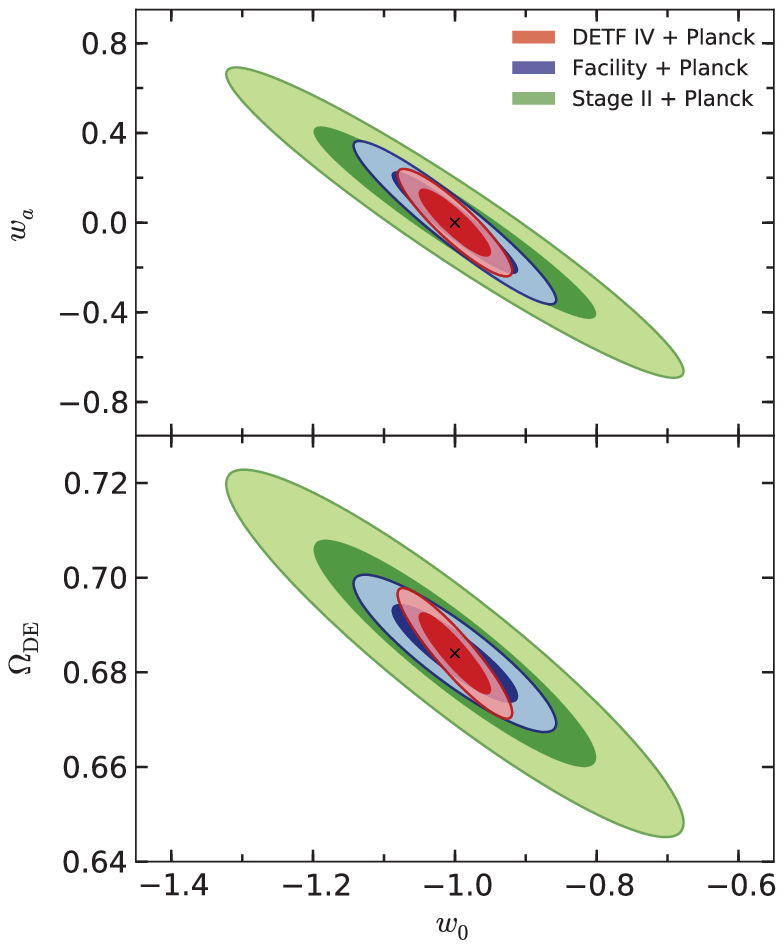}
\caption{{\em Left panel:} Fractional constraints on the dark matter power spectrum $P(k)$ of different planned surveys, combined over the full range over their respective redshift coverage, with 20 bins per decade \citep{Bull2014}. We can see that the "Facility" survey (representative of SKA1 in combined mode) and the "DETF IV" survey (representative of Euclid redshift survey) have comparable measurement power on linear scales ($k \lesssim0.1\,{\rm Mpc}$). {\em Right panel:} Constrains from a Fisher matrix analysis of different combinations of {\em Planck} CMB data with either the "Facility" of "DETF IV" surveys. Again, we see that both combinations have comparable constraining power. Here the dark energy equation of state is parametrised as $w(a)\simeq w_0 + (1-a)w_a$, where $a$ is the scale factor.}
\label{fig:powspec+DE}
\end{figure}

For such an analysis on linear scales, it has been shown that an Intensity Mapping (IM) of the HI fine-structure line emission that does not resolve individual galaxies should provide powerful cosmological constraints \citep{Mao2008}. A recent, detailed forecasting of a joint Planck+SKA1 IM survey \citep{Bull2014}, making full use of the redshift information available, shows that this particular combination of probes should be competitive with e.g. Planck+Euclid redshift survey combinations not only on the standard $\Lambda$CDM model parameters (see Fig.~\ref{fig:LCDM}), but also on dark energy model parameters (see Fig.~\ref{fig:powspec+DE}).

It will also be competitive on constraining the curvature parameter $\Omega_K$, which exhibits a degeneracy with other parameters in CMB data taken alone \citep{Efstathiou1999}, by tightening the constraint by more than a factor of 3. In this analysis, the value of the HI bias has been conservatively marginalised over in each redshift bin. However, a large uncertainty lies in the value of the comoving HI fraction $\Omega_{\rm HI}$ which is poorly known today, and which directly impacts the signal-to-noise of the measured HI emission. This translates into an overall uncertainty of the constraints shown in Figs.~\ref{fig:LCDM},\ref{fig:powspec+DE}.

We should also note that HI emission is strongly contaminated by Galactic (synchrotron, free-free) and extragalactic (free-free, point sources) foreground emissions \citep{Mao2008}. The constraints shown in Figs.~\ref{fig:LCDM},\ref{fig:powspec+DE} are obtained under the assumption that these foreground emissions, due to their smooth spectral emission properties, can be reduced by $10^3$ in amplitude with foreground-cleaning techniques, which represents a significant challenge in itself.

A more direct way to jointly analyse CMB and low-redshift probes is to investigate their cross-correlation, sourced by the correlation of the dark matter fluctuations and the late ISW effect \citep{Sachs1967,Boughn1998,Boughn2004}, possibly using the redshift information \citep{Giannantonio2008,Ho2008}. While this correlation is limited to very large scales and should be therefore cosmic variance limited when using SKA1 data \citep{Raccanelli2012}, it provides a different way of constraining dark energy models, and has a linear (rather than quadratic) dependence on the HI bias. It is therefore an important consistency check of the cosmological model, despite its reduced constraining power. Another way of looking at the correlation of CMB and LSS probes is spatially correlating the extrema of the CMB and LSS on very large scales, possibly assigning the ISW-LSS correlation to the largest supervoids/superclusters around us \citep{Granett2008}. A detailed analysis of these extrema with Planck and SKA1 data will shed a new light on the largest structures around us.

On smaller scales, high frequency SKA measurements of galaxy clusters will provide high-resolution maps of the Sunyaev-Zel'dovich effect \citep{SZ1972, Subhramanian2002} with a very precise subtraction of radio sources (a major contaminant of today's measurements), for up to $10^3$ sources per field of view. Other CMB+SKA synergies include primordial non-gaussianity, high-redshift free-free emission, magnetic fields at cosmological scales are discussed in the other chapter \citep{burigana}.

Finally, CMB fluctuations have a Gaussian distribution at a very accurate level \citep{Planck2013-24}. It is thus possible to estimate, through careful resummation of its trispectrum \citep{Seljak1996,Okamoto2003,Lewis2006,BL2013}, the convergence map of the matter up to $z\sim1080$ \citep{Das2011,VanEngelen2012,Planck2013-17}. The correlation of these CMB convergence maps with the more traditional weak-lensing measurements of background galaxies has been measured recently \citep{Hand2013}. This method, when applied to SKA weak-lensing measurements (see Section~\ref{sec:weak-lensing}) together with CMB data (at large and small scales) will allow to constrain the matter distribution at redshifts unreachable by SKA alone.

The synergies between CMB and SKA data are, in summary, very diverse, and extremely powerful to constrain cosmological parameters, and the first phase SKA1 survey, in combined (single dish plus interferometric) mode, will be competitive with Euclid and LSST for the study of dark energy models.

\section{Galaxy Power Spectrum and Multi-Tracer Method}

In 2020's, the SKA and optical/infrared surveys will perform ultimate observations of large-scale structure of the universe. With huge number of galaxies, the errors in power spectrum of galaxies will be dominated by cosmic variance, rather than shot noise, at cosmological scales. This is especially serious when we try to constrain primordial non-Gaussianity whose effect is stronger at larger scales.

\cite{sel09} proposed a novel method, called multi-tracer method, to defeat cosmic variance using multiple tracers of the dark matter distribution with different bias. Although power spectra of tracers themselves are limited by cosmic variance, the ratio of the power spectra of two tracers, which represents the ralative bias, can evade cosmic variance and is limited only by shot noise. Because the mass and redshift dependences of bias are affected by non-Gaussianity ($f_{\rm NL}$), it can be constrained by the measurements of relative biases.

This multi-tracer method is effective when the bias difference, hence mass difference, is large between tracers and it is critically important to estimate the mass of the dark halo hosting each galaxy. \cite{fer14} considers using different radio galaxy populations (star forming galaxies, starburst galaxies, radio-quiet quasar, FRI and FRII AGN galaxies) as tracers of dark halos of different mass. Although it would not be easy to distinguish these populations observationally, especially between star forming galaxies and starburst, they showed that the SKA continuum survey could ideally reach $f_{\rm NL} \lesssim 1$.

\begin{figure}[tp]
\begin{center}
\includegraphics{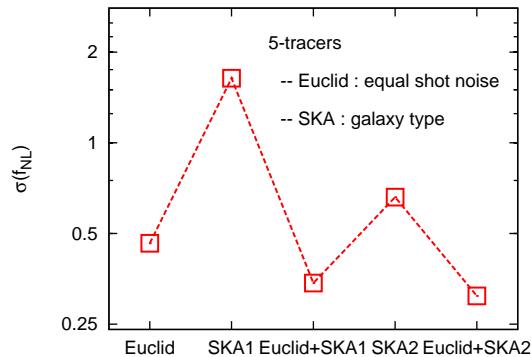}
\caption{Expected constraints on $f_{\rm NL}$ using multi-tracer method assuming observations of the SKA continuum survey, Euclid photometric survey and their combinations \citep{yam14}.}
\label{fig:fNL}
\end{center}
\end{figure}

Another key is the redshift evolution of bias. Because the SKA and optical/IR surveys will have different redshift-distribution of observed galaxies, their combination enhances the power of multi-tracer method. \cite{yam14} studied the potential of combination of the SKA continuum survey and Euclid photometric survey for the constraint on $f_{\rm NL}$. The SKA continuum survey reaches much further than the Euclid photometric survey while the number of galaxies observed by Euclid is larger than that by the SKA at low redshifts, so they are complementary to probe the evolution of bias. Fig. \ref{fig:fNL} shows the expected constraints on $f_{\rm NL}$ from Euclid, SKA1, SKA2 and their combinations. Here, it is assumed that galaxies observed by Euclid have photometric redshift while SKA cannot obtain redshift information. It is seen that the constraint on $f_{\rm NL}$ can reach below unity and it would be possible to approach to $O(0.1)$.

\section{Cluster Cosmology}

The formation of galaxy clusters is seeded by density fluctuations of $10 h^{-1}~{\rm Mpc}$ comoving scale. This is dominated by the gravitational processes and is relatively simple, and thus the halo mass function and its evolution depend strongly on the properties of density fluctuations at the scale. In fact, they have been used as probes for the amplitude of density fluctuations and dark energy. An X-ray satellite eROSITA is expected to be launched in 2016 and will observe about $10^5$ clusters including $10^3$ high-redshift ($z>1$) clusters.

A critical ingredient when we use clusters as a cosmological tool is the estimation of halo mass. There is a scaling relation between halo mass and X-ray observables such as temperature and mass of the intracluster gas. However, because the X-ray observables are sensitive to non-gravitational processes such as radiative cooling \citep{kravtsov,stanek}, the scaling relation has relatively large intrinsic scatter. Thus, it is crucial to calibrate the scaling relation and understand intrinsic scatter in order to do precision cosmology with clusters.

On the other hand, halo mass can be estimated directly by weak lensing of the background galaxies. The optical observations thus far did not have enough sensitivity and, due to the lack of background galaxies, the estimation of halo mass has been successful only for nearby clusters. With a weak lensing survey by the SKA, the estimation of halo mass will become possible for drastically large number of clusters and we will be able to calibrate the scaling relation much more precisely \citep{colafrancesco}.

\section{Summary}

In this chapter, we summarized future survey projects in other wavelengths and showed complementarity and possible synergy with the SKA. In the SKA era with huge number of samples, we need to defeat systematic errors and cosmic variance to advance presicion cosmology.
\begin{itemize}
\item In the case of weak lensing, the instrumental systematics can be reduced by cross-correlating shear signals from the radio and optical surveys. Further, the intrinsic alignment of galaxies, which is difficult to model, can also be probed by their integrated radio polarization.
\item The information on low-redshift universe provided by The galaxy survey of the SKA breaks degeneracies in the estimation of cosmological parameters by the CMB anisotropies. Dark energy can be probed by directly cross-correlating the CMB and galaxy distribution.
\item Euclid and the SKA are complementary in the redshift distribution and so their combination is very effective to study evolution of biases and then constrain primordial non-Gaussianity of density fluctuations.
\item As to cluster cosmology, scaling relation between halo mass and X-ray observables is critical and this can be accurately calibrated by estimation of halo mas with the weak lensing observation of the cluster.
\end{itemize}

K.T. is supported by Grant-in-Aid from the Ministry of Education, Culture, Sports, Science and Technology (MEXT) of Japan, Nos. 24340048 and 26610048. M.L.B. is supported by the an ERC Starting Grant (Grant no. 280127) and by a STFC Advanced/Halliday fellowship. C.B. acknowledge partial support by ASI for {\it Euclid} activities and for the {\it Planck} LFI Activity of Phase E2 (ASI/INAF Agreement 2014-024-R.0). D.Y. is supported by Grant-in-Aid for JSPS Fellows (No.259800).

\bibliographystyle{apj}

\end{document}